\documentclass[aip,reprint,apl]{revtex4-1}
\usepackage{graphicx}

\begin{document}


\title{Measurement of Vacuum Pressure with a Magneto-Optical Trap: a Pressure-Rise Method} 



\author{Rowan W. G. Moore}
\author{Lucie A. Lee}
\author{Elizabeth A. Findlay}
\author{Lara Torralbo-Campo}
\author{Graham D. Bruce}
\author{Donatella Cassettari}
\email[]{dc43@st-andrews.ac.uk}
\affiliation{SUPA School of Physics and Astronomy, University of St Andrews, North Haugh, St Andrews, Fife KY16 9SS, United Kingdom}

\date{\today}

\begin{abstract}
The lifetime of an atom trap is often limited by the presence of residual background gases in the vacuum chamber. This leads to the lifetime being inversely proportional to the pressure. Here we use this dependence to estimate the pressure and to obtain pressure rate-of-rise curves, which are commonly used in vacuum science to evaluate the performance of a system. We observe different rates of pressure increase in response to different levels of outgassing in our system. Therefore we suggest that this is a sensitive method which will find useful applications in cold atom systems, in particular where the inclusion of a standard vacuum gauge is impractical.
\end{abstract}


\maketitle 

There is a trend of making cold atom experiments simpler and more portable in view of taking them outside the laboratory \cite{Knappe06,Sorrentino10,Salim11,Schmidt11,deAngelis11, Barrett14,Farah14,Rushton14}, where they can be used for applications such as precise inertial sensors \cite{Dowling03,Bize05,Budker07,Cronin09,Kitching11}. In a compact apparatus, it is not always practical to include a vacuum gauge, and therefore alternative methods of estimating the background pressure are desirable. Given that pressure is in many cases the dominant factor affecting the lifetime of a trapped sample, the lifetime can in turn be used to estimate the pressure, effectively using the atom trap as a vacuum gauge.\\
\indent In Ref.~\onlinecite{Arpornthip12} this idea was developed into a quantitative method, which we further extend in the present paper by using a Magneto-Optical Trap (MOT) to acquire pressure rate-of-rise curves. These are useful diagnostic tools in vacuum science, and they are taken by turning off the pump after the base pressure of the system has been achieved and monitoring the subsequent pressure increase. The pressure evolution will then indicate whether a real leak is present, in which case the pressure increases linearly with time leading to the determination of the leak size. Or, in absence of real leaks, the pressure as a function of time may reach a plateau, which indicates that an element inside the chamber is outgassing or that a virtual leak (i.e. a small volume of trapped gas) is present. Because the gas released in the chamber in those cases is limited, an equilibrium is reached and the pressure will not increase indefinitely \cite{varian1986basic}. Therefore the pressure-rise method can help establish whether the base pressure in a system is limited by a real leak or by internally-released gas. While pressure--rise curves are commonly measured with a vacuum gauge, in this paper we take the new approach of using the effect of the pressure increase on the MOT. This offers the further advantage that the pressure is measured locally, rather than at a separate location of the vacuum system where the pressure may significantly differ due to limited conductance.\\

\begin{figure}[htb]
		\centering
		\includegraphics{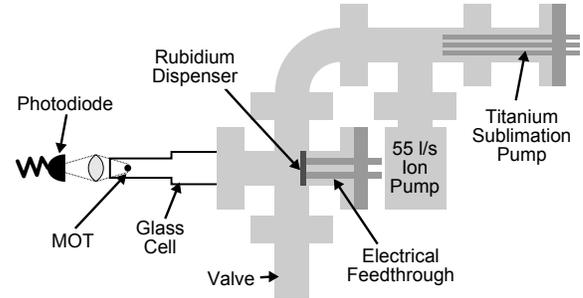}
	\caption[]{Vacuum system: the MOT is created in the glass cell and the trapped atoms are monitored by collecting their fluorescence on a photodiode.}
	\label{fig:exptsetup}
\end{figure}
\indent Our experiment is a vapour cell $^{87}$Rb MOT. Because the MOT selectively loads rubidium atoms, but loses atoms to collisions with untrapped fast rubidium atoms and with other background gases, MOT measurements can be used to extract two distinct contributions to the pressure: that of the rubidium vapour, and that of any other background gas. To separate these contributions, we first characterise our MOT at base pressure (i.e. with pumps on) by using an $N_{\text{eq}}$-$\tau$ plot: we acquire MOT loading curves and measure the equilibrium number of atoms $N_{\text{eq}}$ and the 1/$e$ loading time $\tau$. By repeating these measurements for different levels of rubidium pressure, we gain information on three parameters that characterise the MOT: the trapping cross section, the loss rate due to collisions with non-Rb background gases, and the loss-rate coefficient for the collisions with Rb background. These measurements fully characterise our MOT. To acquire pressure--rise curves, we then turn off the ion pump and monitor the MOT over a period of hours, while the pressure in the system slowly rises. The MOT parameters determined from the initial characterisation are then used to convert these data into quantitative evolutions of the Rb pressure and of the non-Rb pressure. \\
\indent Our six-beam MOT is created in a pyrex cell with 30~mW of optical power and a magnetic field gradient of 18~G/cm. The trapped atoms are detected by collecting fluorescence with a photodiode. The vacuum system is shown in Fig. \ref{fig:exptsetup} and also comprises an isotopically pure $^{87}$Rb dispenser from Alvatec, a 55 L/s ion pump and a titanium sublimation pump. After assembly, the system was baked at $220\,^{\circ}\mathrm{C}$ and a base pressure of the order of 2$\times$10$^{-10}$ Torr was obtained in the ion pump region.\\
%
%
\begin{figure}[htb]
	\centering
		\includegraphics{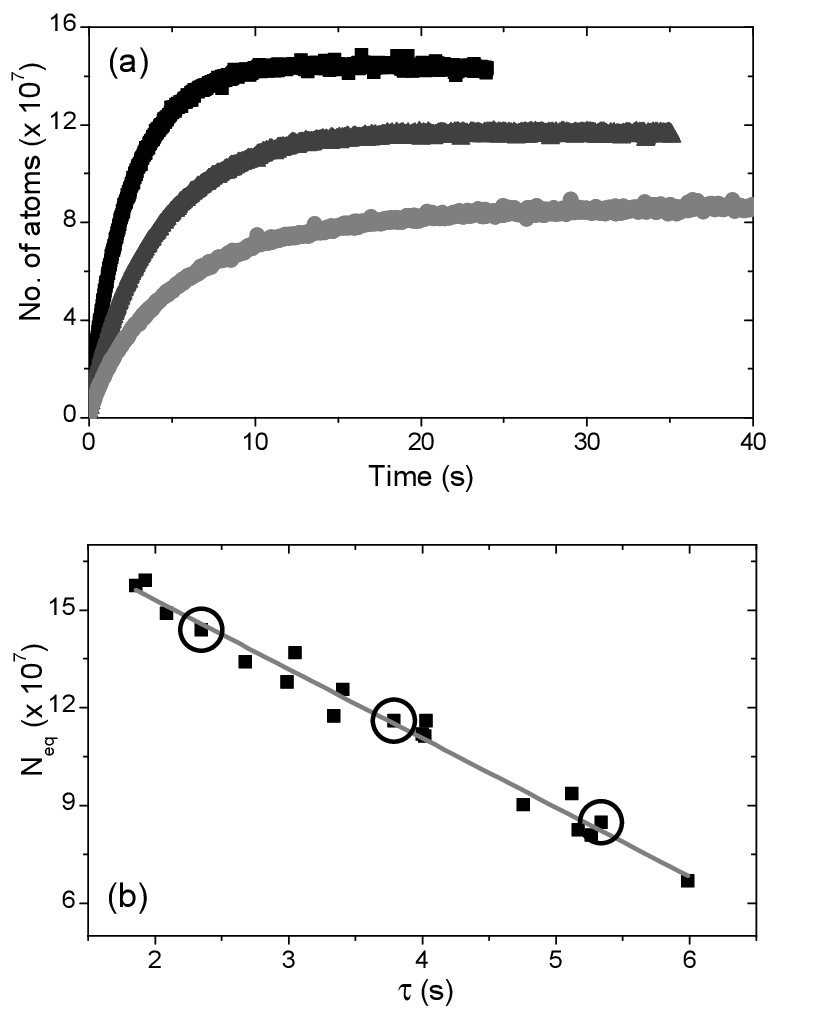}
	\caption[]{Construction of the $N_{\text{eq}}$-$\tau$ plot with
				(a) the MOT loading with different levels of Rb pressure, and
				(b) the resultant $N_{\text{eq}}$-$\tau$ plot where the data shown in (a) have been encircled. The data in (b) are fitted with (\ref{eq:ntau}). 
				}
	\label{fig:n-tau}
\end{figure}\\
\indent For a MOT loaded from background vapour, the MOT dynamics can be well approximated by the following rate equation \cite{Arpornthip12, Steane92}:
\begin{equation}
\label{eq:loading1}
\frac{dN(t)}{dt} = \alpha P_{\text{Rb}} -(\beta P_{\text{Rb}}+\gamma)N(t). 
\end{equation}
\noindent This describes the balance between the rates at which atoms are added to and lost from the trapped population $N$. The first term on the right-hand side is the rate at which atoms are captured; the constant $\alpha$ represents the MOT trapping cross-section while $P_{\text{Rb}}$ is the partial Rb pressure. The second set of terms represents the losses from the trap. The first of these terms, $\beta P_{\text{Rb}}N$, describes losses due to collisions with background Rb atoms. The second term, $\gamma N$, describes losses due to collisions with non-Rb background. \\
\indent The solution of (\ref{eq:loading1}) is
\begin{equation}
\label{eq:loading2}
N(t) = N_{\text{eq}}\left( 1 - \exp\left(-t/\tau\right)\right), 
\end{equation}
\noindent where the equilibrium number of atoms in the MOT is
\begin{equation}
\label{eq:neq}
N_{\text{eq}} = \alpha P_{\text{Rb}}\tau,
\end{equation}
\noindent and the MOT loading time is
\begin{equation}
\label{eq:tau}
\tau = 1/\left(\beta P_{\text{Rb}}+\gamma\right),
\end{equation}
\noindent which coincides with the trap lifetime\cite{Arpornthip12}. Combining (\ref{eq:neq}) and (\ref{eq:tau}) eliminates $P_{\text{Rb}}$, giving
\begin{equation}
\label{eq:ntau}
N_{\text{eq}} = \frac{\alpha}{\beta}(1 - \gamma\tau), 
\end{equation}
\noindent which relates the two easily measurable quantities $N_{\text{eq}}$ and $\tau$. Plotting (\ref{eq:ntau}) experimentally provides the initial characterisation of the MOT. For this purpose, a large amount of rubidium is released into the chamber, after which the Rb source is switched off. A sequence of loading curves is taken as the Rb partial pressure gradually decays, while the non-Rb partial pressure remains constant. This is continued until a data set spanning a sufficiently large range of $N_{\text{eq}}$ and $\tau$ is obtained as shown in Fig. \ref{fig:n-tau}. Fitting these data with (\ref{eq:ntau}) gives $\gamma$ = (0.11 $\pm$ 0.01) s$^{-1}$ and $\alpha/\beta$ = (19.6 $\pm$ 0.3)$\times$10$^{7}$. \\
\begin{figure}[ht]
	\centering
		\includegraphics{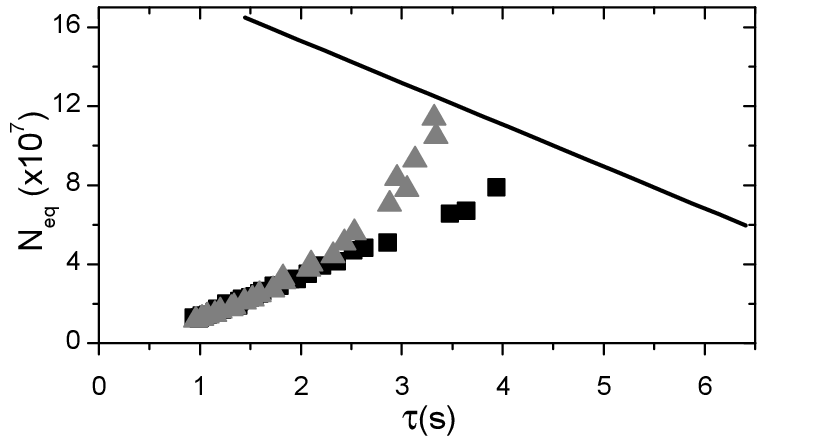}
	\caption[]{The $N_{\text{eq}}$-$\tau$ plots measured with the dispenser current reduced to 4~A and the ion pump turned off. The solid line is the original $N_{\text{eq}}$-$\tau$ plot from Fig. \ref{fig:n-tau}(b). The square and triangular data points are new $N_{\text{eq}}$-$\tau$ plots obtained after the ion pump has been turned off, starting at two different $N_{eq}$ values.}
	\label{fig:n-tau pump off}
\end{figure}

\indent Physically, the value of $\alpha/\beta$ represents the largest MOT that can be obtained in our system, while 1/$\gamma$ is the theoretical upper limit for the loading time as the Rb pressure tends to zero, i.e. the longest possible trap lifetime in our system. This is a useful technique for MOT characterisation that we have previously applied to the study of MOT loading enhanced by Rb pressure modulation \cite{Torralbo14}. In the following this method is further applied to measuring pressure-rise curves in our system to measure vacuum quality and distinguish between different levels of outgassing.\\
%
%
\indent The value of $\gamma$ taken from the linear fit is directly proportional to the non-Rb pressure $P$ in the system. On the assumption that this is mostly due to molecular hydrogen, we use the conversion factor $\gamma/P$ = 4.9$\times$10$^{7}$ Torr$^{-1}$s$^{-1}$ given in Ref. \onlinecite{Arpornthip12}. Combined with $\gamma$ = 0.11 s$^{-1}$ as obtained from Fig. \ref{fig:n-tau}, we estimate a base pressure of 2.2$\times$10$^{-9}$ Torr. This estimate is higher than the value quoted above and the discrepancy can be explained by the limited conductance in our system.\\
 \indent The partial Rb pressure may also be calculated by using the conversion factor $\beta$ = 4.4$\times$10$^{7}$ Torr$^{-1}$s$^{-1}$ as given in Ref. \onlinecite{Arpornthip12}. Thus $\alpha$ can be determined, and hence the rubidium pressure $P_{\text{Rb}}=N_{\text{eq}}/(\alpha\tau)$ (using (\ref{eq:neq})). This pressure varies over the course of the measurements but a typical value is in the 10$^{-9}$ Torr regime.\\
\begin{figure}[t]
	\centering
		\includegraphics{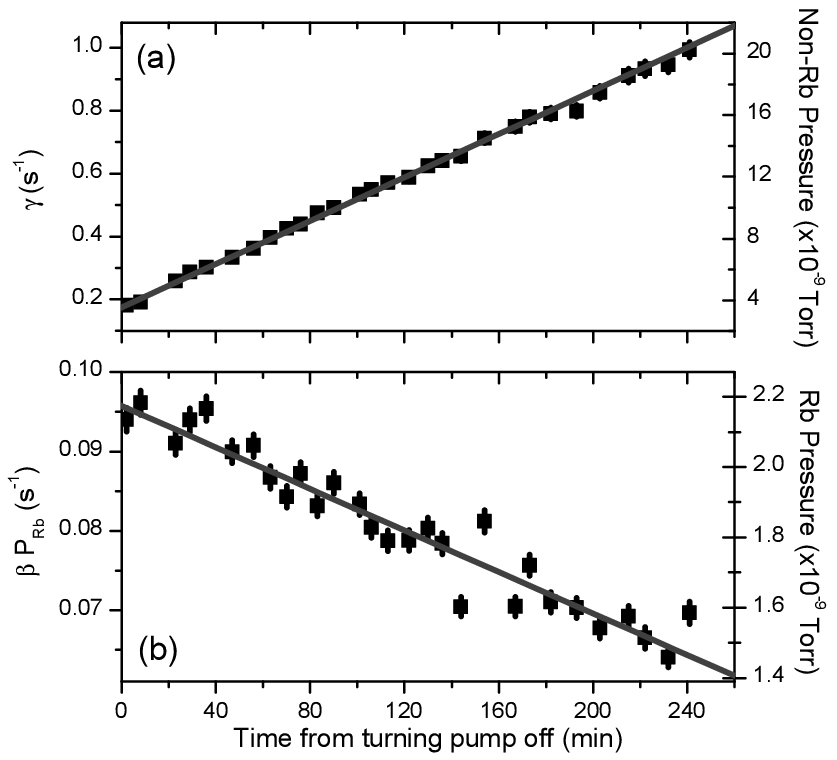}
	\caption[]{Pressure evolution as a function of time after the ion pump has been switched off, extracted from the square data points in Fig. \ref{fig:n-tau pump off}. (a) The non-Rb gases in the system show the expected pressure rise. The linear fit allows the determination of the gas load. (b) The Rb pressure decreases because the dispenser current has been lowered at $t=0$.}
	\label{fig:pressure-rise}
\end{figure}
\indent To obtain pressure--rise curves the procedure is similar to that for MOT characterisation, but with the ion pump switched off at an initial time $t=0$ to allow the non-Rb pressure to rise. Before $t=0$ the dispenser current is set at 5-6 A to trap large numbers of atoms in the MOT, and then is lowered to 4~A at $t=0$ to lower the Rb pressure and prevent overloading the chamber. MOT--loading curves are taken for up to four hours and $N_{\text{eq}}$ is again plotted as a function of $\tau$ as shown in Fig. \ref{fig:n-tau pump off}. Both $N_{eq}$ and $\tau$ decrease over time as the quality of the vacuum deteriorates. This measurement is shown twice, starting from different initial rubidium pressures, and the separate evolutions of $N_{\text{eq}}\left(\tau\right)$ are shown to converge.\\
\indent Using $\alpha/\beta$ obtained from Fig. \ref{fig:n-tau}, the value of $\gamma$ for each time $t$ is calculated by rearranging~(\ref{eq:ntau}):
\begin{equation}
\label{eq:gamma}
\gamma = \frac{1}{\tau}\left(1-\frac{\beta N_{\text{eq}}}{\alpha}\right).
\end{equation}
\begin{figure}[t]
	\centering
		\includegraphics{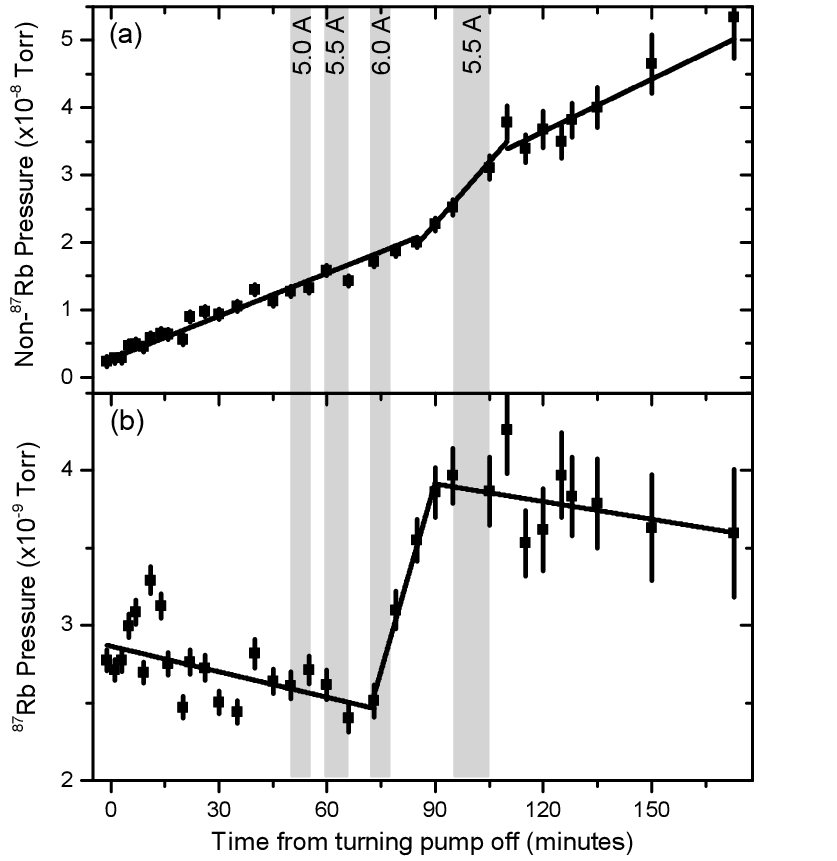}
	\caption[]{Pressure evolution after turning off the ion pump, with pulses applied to the dispenser current. The vertical lines show the duration and current of each pulse. 
	(a) Pressure rise of the non-Rb background showing temporary outgassing. (b) Evolution of the Rb pressure showing a clear increase after the 6~A pulse.}
	\label{fig:dispensers}
\end{figure}
\indent Once again, $\gamma$ is converted to pressure and this pressure is plotted as a function of time in Fig. \ref{fig:pressure-rise}(a), giving a pressure rise curve which is linear in time. From the measured rate of rise and the volume of the chamber ($\sim$1 L), we estimate a gas load of 1.1$\times$10$^{-12}$ Torr L/s. We obtain comparable values for the gas load from both the square and the triangular data points in Fig. \ref{fig:n-tau pump off}, confirming the robustness of this method. We take this gas load as a baseline for the subsequent comparative measurements of outgas rates, and note that this gas load is very low. By comparison, previously baked stainless steel (which constitutes most of the surface in our system) outgasses at a rate of 10$^{-12}$ Torr L/s cm$^{2}$, which corresponds to $>10^{-10}$ Torr L/s for the surface of our system~\cite{VacTec08}. We attribute the observed low rate of pressure rise to the presence of an active titanium layer pumping gas in our system.\\
\indent Our method is capable of discriminating between Rb and non-Rb pressure. The partial Rb pressure is plotted as a function of time in Fig. \ref{fig:pressure-rise}(b), using the conversion outlined above. Due to the reduction in dispenser current at $t=0$, the Rb pressure actually falls over time while the non-Rb pressure is rising. Rubidium pumping is always dominated by high adsorption to the steel walls of the vacuum chamber \cite{Wieman95} and therefore switching off the ion pump has little effect on the Rb pressure.



\indent To test the sensitivity of our method to outgassing, we investigate the effect of repeatedly pulsing the dispenser to higher currents, as shown in Fig. \ref{fig:dispensers}. The current is kept at 4~A between pulses. After the 6~A pulse the Rb pressure clearly increases. However we also see that there is a temporarily increased rise in non-Rb pressure, i.e. an increased gas load. The rate of pressure rise then returns to the pre-pulse level which is comparable to the base gas load shown in Figure \ref{fig:pressure-rise}(a). This is indicative of a temporary outgassing either from within the dispenser, or from a region of the chamber that is being heated up by proximity to the dispenser. By measuring the rate increase in Fig. \ref{fig:dispensers}(a), we estimate a gas load from the 6~A pulse of 6.7$\times$10$^{-12}$~Torr~L/s, which is small and compatible with UHV operation.

\indent In conclusion, we have used an $N_{\text{eq}}$-$\tau$ plot to characterise our MOT, and used the MOT effectively as a vacuum gauge to acquire pressure-rise curves, which quantify outgassing in our vacuum system. The small changes in gas load that we have detected demonstrate the sensitivity of the method. More generally, it should be possible to use this approach to check for leaks in a system and to discriminate between real and virtual leaks.

\indent One particular advantage that this method has over a standard vacuum gauge is that the pressure is measured directly in the MOT region; this is more relevant for cold atoms experiments than the pressure at another point in the system. Moreover, we are able to separate the contribution to the pressure of the rubidium background, which can be monitored for the purpose of characterising the dispenser output. 

\indent We expect this method to find broad applicability to cold atom experiments and to be of particular interest for applications that require a miniaturised vacuum system. \\

\begin{acknowledgments}

\indent R.M. and L.L. contributed equally to this work. We thank Robert Nyman for useful discussions. This research was supported by UK EPSRC, IOP Scotland and the Leverhulme Trust Research Project Grant RPG-2013-074.\\

\end{acknowledgments}



%
%

%


\bibliography{pressure_rise_bib}

\begin{thebibliography}{19}%
\makeatletter
\providecommand \@ifxundefined [1]{%
 \@ifx{#1\undefined}
}%
\providecommand \@ifnum [1]{%
 \ifnum #1\expandafter \@firstoftwo
 \else \expandafter \@secondoftwo
 \fi
}%
\providecommand \@ifx [1]{%
 \ifx #1\expandafter \@firstoftwo
 \else \expandafter \@secondoftwo
 \fi
}%
\providecommand \natexlab [1]{#1}%
\providecommand \enquote  [1]{``#1''}%
\providecommand \bibnamefont  [1]{#1}%
\providecommand \bibfnamefont [1]{#1}%
\providecommand \citenamefont [1]{#1}%
\providecommand \href@noop [0]{\@secondoftwo}%
\providecommand \href [0]{\begingroup \@sanitize@url \@href}%
\providecommand \@href[1]{\@@startlink{#1}\@@href}%
\providecommand \@@href[1]{\endgroup#1\@@endlink}%
\providecommand \@sanitize@url [0]{\catcode `\\12\catcode `\$12\catcode
  `\&12\catcode `\#12\catcode `\^12\catcode `\_12\catcode `\%12\relax}%
\providecommand \@@startlink[1]{}%
\providecommand \@@endlink[0]{}%
\providecommand \url  [0]{\begingroup\@sanitize@url \@url }%
\providecommand \@url [1]{\endgroup\@href {#1}{\urlprefix }}%
\providecommand \urlprefix  [0]{URL }%
\providecommand \Eprint [0]{\href }%
\providecommand \doibase [0]{http://dx.doi.org/}%
\providecommand \selectlanguage [0]{\@gobble}%
\providecommand \bibinfo  [0]{\@secondoftwo}%
\providecommand \bibfield  [0]{\@secondoftwo}%
\providecommand \translation [1]{[#1]}%
\providecommand \BibitemOpen [0]{}%
\providecommand \bibitemStop [0]{}%
\providecommand \bibitemNoStop [0]{.\EOS\space}%
\providecommand \EOS [0]{\spacefactor3000\relax}%
\providecommand \BibitemShut  [1]{\csname bibitem#1\endcsname}%
\let\auto@bib@innerbib\@empty
\bibitem [{\citenamefont {Knappe}\ \emph {et~al.}(2006)\citenamefont {Knappe},
  \citenamefont {Schwindt}, \citenamefont {Gerginov}, \citenamefont {Shah},
  \citenamefont {Liew}, \citenamefont {Moreland}, \citenamefont {Robinson},
  \citenamefont {Hollberg},\ and\ \citenamefont {Kitching}}]{Knappe06}%
  \BibitemOpen
  \bibfield  {author} {\bibinfo {author} {\bibfnamefont {S.}~\bibnamefont
  {Knappe}}, \bibinfo {author} {\bibfnamefont {P.~D.~D.}\ \bibnamefont
  {Schwindt}}, \bibinfo {author} {\bibfnamefont {V.}~\bibnamefont {Gerginov}},
  \bibinfo {author} {\bibfnamefont {V.}~\bibnamefont {Shah}}, \bibinfo {author}
  {\bibfnamefont {L.}~\bibnamefont {Liew}}, \bibinfo {author} {\bibfnamefont
  {J.}~\bibnamefont {Moreland}}, \bibinfo {author} {\bibfnamefont {H.~G.}\
  \bibnamefont {Robinson}}, \bibinfo {author} {\bibfnamefont {L.}~\bibnamefont
  {Hollberg}}, \ and\ \bibinfo {author} {\bibfnamefont {J.}~\bibnamefont
  {Kitching}},\ }\href@noop {} {\bibfield  {journal} {\bibinfo  {journal} {J.
  Opt. A: Pure Appl. Opt.}\ }\textbf {\bibinfo {volume} {8}},\ \bibinfo {pages}
  {S318} (\bibinfo {year} {2006})}\BibitemShut {NoStop}%
\bibitem [{\citenamefont {Sorrentino}\ \emph {et~al.}(2010)\citenamefont
  {Sorrentino}, \citenamefont {Bongs}, \citenamefont {Bouyer}, \citenamefont
  {Cacciapuoti}, \citenamefont {de~Angelis}, \citenamefont {Dittus},
  \citenamefont {Ertmer}, \citenamefont {Giorgini}, \citenamefont {Hartwig},
  \citenamefont {Hauth}, \citenamefont {Herrmann}, \citenamefont {Inguscio},
  \citenamefont {Kajari}, \citenamefont {K\"onemann}, \citenamefont
  {L\"ammerzahl}, \citenamefont {Landragin}, \citenamefont {Modugno},
  \citenamefont {Pereira~dos Santos}, \citenamefont {Peters}, \citenamefont
  {Prevedelli}, \citenamefont {Rasel}, \citenamefont {Schleich}, \citenamefont
  {Schmidt}, \citenamefont {Senger}, \citenamefont {Sengstock}, \citenamefont
  {Stern}, \citenamefont {Tino},\ and\ \citenamefont {Walser}}]{Sorrentino10}%
  \BibitemOpen
  \bibfield  {author} {\bibinfo {author} {\bibfnamefont {F.}~\bibnamefont
  {Sorrentino}}, \bibinfo {author} {\bibfnamefont {K.}~\bibnamefont {Bongs}},
  \bibinfo {author} {\bibfnamefont {P.}~\bibnamefont {Bouyer}}, \bibinfo
  {author} {\bibfnamefont {L.}~\bibnamefont {Cacciapuoti}}, \bibinfo {author}
  {\bibfnamefont {M.}~\bibnamefont {de~Angelis}}, \bibinfo {author}
  {\bibfnamefont {H.}~\bibnamefont {Dittus}}, \bibinfo {author} {\bibfnamefont
  {W.}~\bibnamefont {Ertmer}}, \bibinfo {author} {\bibfnamefont
  {A.}~\bibnamefont {Giorgini}}, \bibinfo {author} {\bibfnamefont
  {J.}~\bibnamefont {Hartwig}}, \bibinfo {author} {\bibfnamefont
  {M.}~\bibnamefont {Hauth}}, \bibinfo {author} {\bibfnamefont
  {S.}~\bibnamefont {Herrmann}}, \bibinfo {author} {\bibfnamefont
  {M.}~\bibnamefont {Inguscio}}, \bibinfo {author} {\bibfnamefont
  {E.}~\bibnamefont {Kajari}}, \bibinfo {author} {\bibfnamefont {T.~T.}\
  \bibnamefont {K\"onemann}}, \bibinfo {author} {\bibfnamefont
  {C.}~\bibnamefont {L\"ammerzahl}}, \bibinfo {author} {\bibfnamefont
  {A.}~\bibnamefont {Landragin}}, \bibinfo {author} {\bibfnamefont
  {G.}~\bibnamefont {Modugno}}, \bibinfo {author} {\bibfnamefont
  {F.}~\bibnamefont {Pereira~dos Santos}}, \bibinfo {author} {\bibfnamefont
  {A.}~\bibnamefont {Peters}}, \bibinfo {author} {\bibfnamefont
  {M.}~\bibnamefont {Prevedelli}}, \bibinfo {author} {\bibfnamefont
  {E.}~\bibnamefont {Rasel}}, \bibinfo {author} {\bibfnamefont
  {W.}~\bibnamefont {Schleich}}, \bibinfo {author} {\bibfnamefont
  {M.}~\bibnamefont {Schmidt}}, \bibinfo {author} {\bibfnamefont
  {A.}~\bibnamefont {Senger}}, \bibinfo {author} {\bibfnamefont
  {K.}~\bibnamefont {Sengstock}}, \bibinfo {author} {\bibfnamefont
  {G.}~\bibnamefont {Stern}}, \bibinfo {author} {\bibfnamefont
  {G.}~\bibnamefont {Tino}}, \ and\ \bibinfo {author} {\bibfnamefont
  {R.}~\bibnamefont {Walser}},\ }\href@noop {} {\bibfield  {journal} {\bibinfo
  {journal} {Microgravity Sci. Technol.}\ }\textbf {\bibinfo {volume} {22}},\
  \bibinfo {pages} {551} (\bibinfo {year} {2010})}\BibitemShut {NoStop}%
\bibitem [{\citenamefont {Salim}\ \emph {et~al.}(2011)\citenamefont {Salim},
  \citenamefont {DeNatale}, \citenamefont {Farkas}, \citenamefont {Hudek},
  \citenamefont {McBride}, \citenamefont {Michalchuk}, \citenamefont
  {Mihailovich},\ and\ \citenamefont {Anderson}}]{Salim11}%
  \BibitemOpen
  \bibfield  {author} {\bibinfo {author} {\bibfnamefont {E.~A.}\ \bibnamefont
  {Salim}}, \bibinfo {author} {\bibfnamefont {J.}~\bibnamefont {DeNatale}},
  \bibinfo {author} {\bibfnamefont {D.~M.}\ \bibnamefont {Farkas}}, \bibinfo
  {author} {\bibfnamefont {K.~M.}\ \bibnamefont {Hudek}}, \bibinfo {author}
  {\bibfnamefont {S.~E.}\ \bibnamefont {McBride}}, \bibinfo {author}
  {\bibfnamefont {J.}~\bibnamefont {Michalchuk}}, \bibinfo {author}
  {\bibfnamefont {R.}~\bibnamefont {Mihailovich}}, \ and\ \bibinfo {author}
  {\bibfnamefont {D.~Z.}\ \bibnamefont {Anderson}},\ }\href@noop {} {\bibfield
  {journal} {\bibinfo  {journal} {Quantum Inf. Process.}\ }\textbf {\bibinfo
  {volume} {10}},\ \bibinfo {pages} {975} (\bibinfo {year} {2011})}\BibitemShut
  {NoStop}%
\bibitem [{\citenamefont {Schmidt}\ \emph {et~al.}(2011)\citenamefont
  {Schmidt}, \citenamefont {Senger}, \citenamefont {Hauth}, \citenamefont
  {Freier}, \citenamefont {Schkolnik},\ and\ \citenamefont
  {Peters}}]{Schmidt11}%
  \BibitemOpen
  \bibfield  {author} {\bibinfo {author} {\bibfnamefont {M.}~\bibnamefont
  {Schmidt}}, \bibinfo {author} {\bibfnamefont {A.}~\bibnamefont {Senger}},
  \bibinfo {author} {\bibfnamefont {M.}~\bibnamefont {Hauth}}, \bibinfo
  {author} {\bibfnamefont {C.}~\bibnamefont {Freier}}, \bibinfo {author}
  {\bibfnamefont {V.}~\bibnamefont {Schkolnik}}, \ and\ \bibinfo {author}
  {\bibfnamefont {A.}~\bibnamefont {Peters}},\ }\href@noop {} {\bibfield
  {journal} {\bibinfo  {journal} {Gyrosc. Navig.}\ }\textbf {\bibinfo {volume}
  {2}},\ \bibinfo {pages} {170} (\bibinfo {year} {2011})}\BibitemShut {NoStop}%
\bibitem [{\citenamefont {{de Angelis}}\ \emph {et~al.}(2011)\citenamefont {{de
  Angelis}}, \citenamefont {Angonin}, \citenamefont {Beaufils}, \citenamefont
  {Becker}, \citenamefont {Bertoldi}, \citenamefont {Bongs}, \citenamefont
  {Bourdel}, \citenamefont {Bouyer}, \citenamefont {Boyer}, \citenamefont
  {D\"{o}rscher}, \citenamefont {Duncker}, \citenamefont {Ertmer},
  \citenamefont {Fernholz}, \citenamefont {Fromhold}, \citenamefont {Herr},
  \citenamefont {Kr\"{u}ger}, \citenamefont {K\"{u}rbis}, \citenamefont
  {Mellor}, \citenamefont {Santos}, \citenamefont {Peters}, \citenamefont
  {Poli}, \citenamefont {Popp}, \citenamefont {Prevedelli}, \citenamefont
  {Rasel}, \citenamefont {Rudolph}, \citenamefont {Schreck}, \citenamefont
  {Sengstock}, \citenamefont {Sorrentino}, \citenamefont {Stellmer},
  \citenamefont {Tino}, \citenamefont {Valenzuela}, \citenamefont {Wendrich},
  \citenamefont {Wicht}, \citenamefont {Windpassinger},\ and\ \citenamefont
  {Wolf}}]{deAngelis11}%
  \BibitemOpen
  \bibfield  {author} {\bibinfo {author} {\bibfnamefont {M.}~\bibnamefont {{de
  Angelis}}}, \bibinfo {author} {\bibfnamefont {M.}~\bibnamefont {Angonin}},
  \bibinfo {author} {\bibfnamefont {Q.}~\bibnamefont {Beaufils}}, \bibinfo
  {author} {\bibfnamefont {C.}~\bibnamefont {Becker}}, \bibinfo {author}
  {\bibfnamefont {A.}~\bibnamefont {Bertoldi}}, \bibinfo {author}
  {\bibfnamefont {K.}~\bibnamefont {Bongs}}, \bibinfo {author} {\bibfnamefont
  {T.}~\bibnamefont {Bourdel}}, \bibinfo {author} {\bibfnamefont
  {P.}~\bibnamefont {Bouyer}}, \bibinfo {author} {\bibfnamefont
  {V.}~\bibnamefont {Boyer}}, \bibinfo {author} {\bibfnamefont
  {S.}~\bibnamefont {D\"{o}rscher}}, \bibinfo {author} {\bibfnamefont
  {H.}~\bibnamefont {Duncker}}, \bibinfo {author} {\bibfnamefont
  {W.}~\bibnamefont {Ertmer}}, \bibinfo {author} {\bibfnamefont
  {T.}~\bibnamefont {Fernholz}}, \bibinfo {author} {\bibfnamefont {T.~M.}\
  \bibnamefont {Fromhold}}, \bibinfo {author} {\bibfnamefont {W.}~\bibnamefont
  {Herr}}, \bibinfo {author} {\bibfnamefont {P.}~\bibnamefont {Kr\"{u}ger}},
  \bibinfo {author} {\bibfnamefont {C.}~\bibnamefont {K\"{u}rbis}}, \bibinfo
  {author} {\bibfnamefont {C.}~\bibnamefont {Mellor}}, \bibinfo {author}
  {\bibfnamefont {F.~P.~D.}\ \bibnamefont {Santos}}, \bibinfo {author}
  {\bibfnamefont {A.}~\bibnamefont {Peters}}, \bibinfo {author} {\bibfnamefont
  {N.}~\bibnamefont {Poli}}, \bibinfo {author} {\bibfnamefont {M.}~\bibnamefont
  {Popp}}, \bibinfo {author} {\bibfnamefont {M.}~\bibnamefont {Prevedelli}},
  \bibinfo {author} {\bibfnamefont {E.}~\bibnamefont {Rasel}}, \bibinfo
  {author} {\bibfnamefont {J.}~\bibnamefont {Rudolph}}, \bibinfo {author}
  {\bibfnamefont {F.}~\bibnamefont {Schreck}}, \bibinfo {author} {\bibfnamefont
  {K.}~\bibnamefont {Sengstock}}, \bibinfo {author} {\bibfnamefont
  {F.}~\bibnamefont {Sorrentino}}, \bibinfo {author} {\bibfnamefont
  {S.}~\bibnamefont {Stellmer}}, \bibinfo {author} {\bibfnamefont
  {G.}~\bibnamefont {Tino}}, \bibinfo {author} {\bibfnamefont {T.}~\bibnamefont
  {Valenzuela}}, \bibinfo {author} {\bibfnamefont {T.}~\bibnamefont
  {Wendrich}}, \bibinfo {author} {\bibfnamefont {A.}~\bibnamefont {Wicht}},
  \bibinfo {author} {\bibfnamefont {P.}~\bibnamefont {Windpassinger}}, \ and\
  \bibinfo {author} {\bibfnamefont {P.}~\bibnamefont {Wolf}},\ }\href@noop {}
  {\bibfield  {journal} {\bibinfo  {journal} {Procedia Comput. Sci.}\ }\textbf
  {\bibinfo {volume} {7}},\ \bibinfo {pages} {334 } (\bibinfo {year}
  {2011})}\BibitemShut {NoStop}%
\bibitem [{\citenamefont {Barrett}\ \emph {et~al.}(2014)\citenamefont
  {Barrett}, \citenamefont {Gominet}, \citenamefont {Cantin}, \citenamefont
  {Antoni-Micollier}, \citenamefont {Bertoldi}, \citenamefont {Battelier},
  \citenamefont {Bouyer}, \citenamefont {Lautier},\ and\ \citenamefont
  {Landragin}}]{Barrett14}%
  \BibitemOpen
  \bibfield  {author} {\bibinfo {author} {\bibfnamefont {B.}~\bibnamefont
  {Barrett}}, \bibinfo {author} {\bibfnamefont {P.-A.}\ \bibnamefont
  {Gominet}}, \bibinfo {author} {\bibfnamefont {E.}~\bibnamefont {Cantin}},
  \bibinfo {author} {\bibfnamefont {L.}~\bibnamefont {Antoni-Micollier}},
  \bibinfo {author} {\bibfnamefont {A.}~\bibnamefont {Bertoldi}}, \bibinfo
  {author} {\bibfnamefont {B.}~\bibnamefont {Battelier}}, \bibinfo {author}
  {\bibfnamefont {P.}~\bibnamefont {Bouyer}}, \bibinfo {author} {\bibfnamefont
  {J.}~\bibnamefont {Lautier}}, \ and\ \bibinfo {author} {\bibfnamefont
  {A.}~\bibnamefont {Landragin}},\ }in\ \href@noop {} {\emph {\bibinfo
  {booktitle} {Proceedings of the International School of Physics ``Enrico
  Fermi''}}},\ Vol.\ \bibinfo {volume} {188: Atom Interferometry}\ (\bibinfo
  {publisher} {IOS Press},\ \bibinfo {year} {2014})\ pp.\ \bibinfo {pages}
  {493--555}\BibitemShut {NoStop}%
\bibitem [{\citenamefont {Farah}\ \emph {et~al.}(2014)\citenamefont {Farah},
  \citenamefont {Guerlin}, \citenamefont {Landragin}, \citenamefont {Bouyer},
  \citenamefont {Gaffet}, \citenamefont {Pereira Dos~Santos},\ and\
  \citenamefont {Merlet}}]{Farah14}%
  \BibitemOpen
  \bibfield  {author} {\bibinfo {author} {\bibfnamefont {T.}~\bibnamefont
  {Farah}}, \bibinfo {author} {\bibfnamefont {C.}~\bibnamefont {Guerlin}},
  \bibinfo {author} {\bibfnamefont {A.}~\bibnamefont {Landragin}}, \bibinfo
  {author} {\bibfnamefont {P.}~\bibnamefont {Bouyer}}, \bibinfo {author}
  {\bibfnamefont {S.}~\bibnamefont {Gaffet}}, \bibinfo {author} {\bibfnamefont
  {F.}~\bibnamefont {Pereira Dos~Santos}}, \ and\ \bibinfo {author}
  {\bibfnamefont {S.}~\bibnamefont {Merlet}},\ }\href@noop {} {\bibfield
  {journal} {\bibinfo  {journal} {Gyrosc. Navig.}\ }\textbf {\bibinfo {volume}
  {5}},\ \bibinfo {pages} {266} (\bibinfo {year} {2014})}\BibitemShut {NoStop}%
\bibitem [{\citenamefont {Rushton}, \citenamefont {Aldous},\ and\ \citenamefont
  {Himsworth}(2014)}]{Rushton14}%
  \BibitemOpen
  \bibfield  {author} {\bibinfo {author} {\bibfnamefont {J.}~\bibnamefont
  {Rushton}}, \bibinfo {author} {\bibfnamefont {M.}~\bibnamefont {Aldous}}, \
  and\ \bibinfo {author} {\bibfnamefont {M.}~\bibnamefont {Himsworth}},\
  }\href@noop {} {\enquote {\bibinfo {title} {{The Feasibility of a Fully
  Miniaturized Magneto-Optical Trap for Portable Ultracold Quantum
  Technology}},}\ } (\bibinfo {year} {2014}),\ \bibinfo {note}
  {\url{http://arxiv.org/abs/1405.3148}}\BibitemShut {NoStop}%
\bibitem [{\citenamefont {Dowling}\ and\ \citenamefont
  {Milburn}(2003)}]{Dowling03}%
  \BibitemOpen
  \bibfield  {author} {\bibinfo {author} {\bibfnamefont {J.~P.}\ \bibnamefont
  {Dowling}}\ and\ \bibinfo {author} {\bibfnamefont {G.~J.}\ \bibnamefont
  {Milburn}},\ }\href@noop {} {\bibfield  {journal} {\bibinfo  {journal} {Phil.
  Trans. R. Soc. Lond. A}\ }\textbf {\bibinfo {volume} {361}},\ \bibinfo
  {pages} {1655} (\bibinfo {year} {2003})}\BibitemShut {NoStop}%
\bibitem [{\citenamefont {Bize}\ \emph {et~al.}(2005)\citenamefont {Bize},
  \citenamefont {Laurent}, \citenamefont {Abgrall}, \citenamefont {Marion},
  \citenamefont {Maksimovic}, \citenamefont {Cacciapuoti}, \citenamefont
  {Gr\"unert}, \citenamefont {Vian}, \citenamefont {{Pereira dos Santos}},
  \citenamefont {Rosenbusch}, \citenamefont {Lemonde}, \citenamefont
  {Santarelli}, \citenamefont {Wolf}, \citenamefont {Clairon}, \citenamefont
  {Luiten}, \citenamefont {Tobar},\ and\ \citenamefont {Salomon}}]{Bize05}%
  \BibitemOpen
  \bibfield  {author} {\bibinfo {author} {\bibfnamefont {S.}~\bibnamefont
  {Bize}}, \bibinfo {author} {\bibfnamefont {P.}~\bibnamefont {Laurent}},
  \bibinfo {author} {\bibfnamefont {M.}~\bibnamefont {Abgrall}}, \bibinfo
  {author} {\bibfnamefont {H.}~\bibnamefont {Marion}}, \bibinfo {author}
  {\bibfnamefont {I.}~\bibnamefont {Maksimovic}}, \bibinfo {author}
  {\bibfnamefont {L.}~\bibnamefont {Cacciapuoti}}, \bibinfo {author}
  {\bibfnamefont {J.}~\bibnamefont {Gr\"unert}}, \bibinfo {author}
  {\bibfnamefont {C.}~\bibnamefont {Vian}}, \bibinfo {author} {\bibfnamefont
  {F.}~\bibnamefont {{Pereira dos Santos}}}, \bibinfo {author} {\bibfnamefont
  {P.}~\bibnamefont {Rosenbusch}}, \bibinfo {author} {\bibfnamefont
  {P.}~\bibnamefont {Lemonde}}, \bibinfo {author} {\bibfnamefont
  {G.}~\bibnamefont {Santarelli}}, \bibinfo {author} {\bibfnamefont
  {P.}~\bibnamefont {Wolf}}, \bibinfo {author} {\bibfnamefont {A.}~\bibnamefont
  {Clairon}}, \bibinfo {author} {\bibfnamefont {A.}~\bibnamefont {Luiten}},
  \bibinfo {author} {\bibfnamefont {M.}~\bibnamefont {Tobar}}, \ and\ \bibinfo
  {author} {\bibfnamefont {C.}~\bibnamefont {Salomon}},\ }\href@noop {}
  {\bibfield  {journal} {\bibinfo  {journal} {J. Phys. B: At. Mol. Opt. Phys.}\
  }\textbf {\bibinfo {volume} {38}},\ \bibinfo {pages} {S449} (\bibinfo {year}
  {2005})}\BibitemShut {NoStop}%
\bibitem [{\citenamefont {Budker}\ and\ \citenamefont
  {Romalis}(2007)}]{Budker07}%
  \BibitemOpen
  \bibfield  {author} {\bibinfo {author} {\bibfnamefont {D.}~\bibnamefont
  {Budker}}\ and\ \bibinfo {author} {\bibfnamefont {M.}~\bibnamefont
  {Romalis}},\ }\href@noop {} {\bibfield  {journal} {\bibinfo  {journal} {Nat.
  Phys.}\ }\textbf {\bibinfo {volume} {3}},\ \bibinfo {pages} {227} (\bibinfo
  {year} {2007})}\BibitemShut {NoStop}%
\bibitem [{\citenamefont {Cronin}, \citenamefont {Schmiedmayer},\ and\
  \citenamefont {Pritchard}(2009)}]{Cronin09}%
  \BibitemOpen
  \bibfield  {author} {\bibinfo {author} {\bibfnamefont {A.~D.}\ \bibnamefont
  {Cronin}}, \bibinfo {author} {\bibfnamefont {J.}~\bibnamefont
  {Schmiedmayer}}, \ and\ \bibinfo {author} {\bibfnamefont {D.~E.}\
  \bibnamefont {Pritchard}},\ }\href@noop {} {\bibfield  {journal} {\bibinfo
  {journal} {Rev. Mod. Phys.}\ }\textbf {\bibinfo {volume} {81}},\ \bibinfo
  {pages} {1051} (\bibinfo {year} {2009})}\BibitemShut {NoStop}%
\bibitem [{\citenamefont {Kitching}, \citenamefont {Knappe},\ and\
  \citenamefont {Donley}(2011)}]{Kitching11}%
  \BibitemOpen
  \bibfield  {author} {\bibinfo {author} {\bibfnamefont {J.}~\bibnamefont
  {Kitching}}, \bibinfo {author} {\bibfnamefont {S.}~\bibnamefont {Knappe}}, \
  and\ \bibinfo {author} {\bibfnamefont {E.~A.}\ \bibnamefont {Donley}},\
  }\href@noop {} {\bibfield  {journal} {\bibinfo  {journal} {IEEE Sensors
  Journal}\ }\textbf {\bibinfo {volume} {11}},\ \bibinfo {pages} {1749}
  (\bibinfo {year} {2011})}\BibitemShut {NoStop}%
\bibitem [{\citenamefont {{Arpornthip}}, \citenamefont {{Sackett}},\ and\
  \citenamefont {{Hughes}}(2012)}]{Arpornthip12}%
  \BibitemOpen
  \bibfield  {author} {\bibinfo {author} {\bibfnamefont {T.}~\bibnamefont
  {{Arpornthip}}}, \bibinfo {author} {\bibfnamefont {C.~A.}\ \bibnamefont
  {{Sackett}}}, \ and\ \bibinfo {author} {\bibfnamefont {K.~J.}\ \bibnamefont
  {{Hughes}}},\ }\href@noop {} {\bibfield  {journal} {\bibinfo  {journal}
  {Phys. Rev. A}\ }\textbf {\bibinfo {volume} {85}},\ \bibinfo {pages} {033420}
  (\bibinfo {year} {2012})}\BibitemShut {NoStop}%
\bibitem [{var(1986)}]{varian1986basic}%
  \BibitemOpen
  \href@noop {} {\emph {\bibinfo {title} {Basic Vacuum Practice}}}\ (\bibinfo
  {publisher} {Varian Associates, Incorporated},\ \bibinfo {year}
  {1986})\BibitemShut {NoStop}%
\bibitem [{\citenamefont {Steane}, \citenamefont {Chowdhury},\ and\
  \citenamefont {Foot}(1992)}]{Steane92}%
  \BibitemOpen
  \bibfield  {author} {\bibinfo {author} {\bibfnamefont {A.~M.}\ \bibnamefont
  {Steane}}, \bibinfo {author} {\bibfnamefont {M.}~\bibnamefont {Chowdhury}}, \
  and\ \bibinfo {author} {\bibfnamefont {C.~J.}\ \bibnamefont {Foot}},\
  }\href@noop {} {\bibfield  {journal} {\bibinfo  {journal} {J. Opt. Soc. Am.
  B}\ }\textbf {\bibinfo {volume} {9}},\ \bibinfo {pages} {2142} (\bibinfo
  {year} {1992})}\BibitemShut {NoStop}%
\bibitem [{\citenamefont {Torralbo-Campo}\ \emph {et~al.}(2014)\citenamefont
  {Torralbo-Campo}, \citenamefont {Bruce}, \citenamefont {Smirne},\ and\
  \citenamefont {Cassettari}}]{Torralbo14}%
  \BibitemOpen
  \bibfield  {author} {\bibinfo {author} {\bibfnamefont {L.}~\bibnamefont
  {Torralbo-Campo}}, \bibinfo {author} {\bibfnamefont {G.~D.}\ \bibnamefont
  {Bruce}}, \bibinfo {author} {\bibfnamefont {G.}~\bibnamefont {Smirne}}, \
  and\ \bibinfo {author} {\bibfnamefont {D.}~\bibnamefont {Cassettari}},\
  }\href@noop {} {\enquote {\bibinfo {title} {Light-induced atomic desorption
  in a compact system for ultracold atoms},}\ } (\bibinfo {year} {2014}),\
  \bibinfo {note} {submitted to Phys. Rev. A
  \url{http://arxiv.org/abs/1312.6442}}\BibitemShut {NoStop}%
\bibitem [{\citenamefont {{Yoshimura}}(2008)}]{VacTec08}%
  \BibitemOpen
  \bibfield  {author} {\bibinfo {author} {\bibfnamefont {N.}~\bibnamefont
  {{Yoshimura}}},\ }\href@noop {} {\emph {\bibinfo {title} {Vacuum
  Technology}}}\ (\bibinfo  {publisher} {Chapman and Hall},\ \bibinfo {address}
  {Tokyo},\ \bibinfo {year} {2008})\BibitemShut {NoStop}%
\bibitem [{\citenamefont {{Wieman}}, \citenamefont {{Flowers}},\ and\
  \citenamefont {{Gilbert}}(1995)}]{Wieman95}%
  \BibitemOpen
  \bibfield  {author} {\bibinfo {author} {\bibfnamefont {C.}~\bibnamefont
  {{Wieman}}}, \bibinfo {author} {\bibfnamefont {G.}~\bibnamefont {{Flowers}}},
  \ and\ \bibinfo {author} {\bibfnamefont {S.}~\bibnamefont {{Gilbert}}},\
  }\href@noop {} {\bibfield  {journal} {\bibinfo  {journal} {Am. J. Phys.}\
  }\textbf {\bibinfo {volume} {63}},\ \bibinfo {pages} {317} (\bibinfo {year}
  {1995})}\BibitemShut {NoStop}%
\end{thebibliography}%

\end{document}